\documentclass{PoS}
\title{$B_s$ Decays at the Tevatron}

\ShortTitle{$B_s$ Decays at the Tevatron}

\author{\speaker{Gavril Giurgiu}\thanks{on behalf of the CDF and D0 Collaborations}\\
        Johns Hopkins University\\
        E-mail: \email{ggiurgiu@jhu.edu}}

\abstract{
	We present measurements of the branching ratio and of the polarization 
	amplitudes in charmless $B_s \rightarrow \phi \phi$ decays using data 
	corresponding to 2.9~fb$^{-1}$ of integrated luminosity, collected 
	by the CDF experiment at the Tevatron. The branching ratio in 
	$B_s \rightarrow \phi \phi$ decays is measured relative to the normalization 
	mode $B_s \rightarrow J/\psi \phi$ to be 
	${\cal B}(B_s \rightarrow \phi \phi) / {\cal B}(B_s \rightarrow J/\psi \phi) = [1.78 \pm 0.14 (stat) \pm 0.20 (syst)] \times 10^{-2}$.
	Using the experimental value of ${\cal B}(B_s \rightarrow J/\psi \phi)$ we determine the 
	$B_s \rightarrow \phi \phi$ branching ratio 
	\begin{center}
		${\cal B}(B_s \rightarrow \phi \phi) = 2.40 \pm 0.21 (stat) \pm 0.27 (syst) \pm 0.82 (BR)] \times 10^{-5}.$
	\end{center}
	The polarization fractions are measured for the first time in this analysis 
	and found to be:
	\begin{center}
		$|A_0|^2 = 0.348 \pm 0.041 (stat) \pm 0.021 (syst)$ \\	
		$|A_{\parallel}|^2 = 0.287 \pm 0.043 (stat) \pm 0.011 (syst)$ \\
		$|A_{\perp}|^2 = 0.365 \pm 0.044 (stat) \pm 0.027 (syst)$.
	\end{center}		
	   
}

\FullConference{Flavor Physics and CP Violation - FPCP 2010\\
		May 25-29, 2010\\
		Turin, Italy}

\begin{document}

\section{Introduction}

$B_s$ mesons were initially studied by the LEP experiments and then by the 
CLEO experiment at $\Upsilon(5S)$. More recently the KEKB accelerator has 
been running at $\Upsilon(5S)$ resonance as well, enabling the Belle experiment to 
do $B_s$ physics. The largest $B_s$ samples, however, are collected by the 
CDF and D0 experiments at the Fermilab Tevatron. To date, the Tevatron 
has delivered about 8~fb$^{-1}$ of integrated luminosity while each of the 
two Tevatron experiments has recorded close to 7~fb$^{-1}$ of integrated 
luminosity on tape. 
The most recent Tevatron results in $B_s$ physics include studies of rare 
decays~\cite{rare_decays} like $B_s \rightarrow \mu \mu$, $B_s \rightarrow e \mu$ 
or $B_s \rightarrow \phi \mu \mu$, 
CP violation in $B_s \rightarrow J/\psi \phi$ decays~\cite{CDF_beta_s, D0_beta_s} 
and CP violation in inclusive semileptonic $B$ decays~\cite{D0_di_muon}. 

In this paper we focus on studies of charmless $B_s \rightarrow \phi \phi$ decays, 
performed by the CDF experiment at Fermilab. We present measurements of the branching 
ratio~\cite{CDF_br} and of the polarization fractions~\cite{CDF_pol} in these decays using 
data corresponding to 2.9~fb$^{-1}$ of integrated luminosity. 

Charmless $B_s$ decays are still to be fully understood. They offer the possibility 
to test our current theoretical understanding and represent promising ways to search 
for physics beyond the Standard Model (SM). 
The $B_s \rightarrow \phi \phi$ decay is part of the so called $B \rightarrow VV$ family 
in which the initial state $B$-meson is a pseudo-scalar 
(spin 0) and the final state $VV$ contains two vector 
mesons (spin 1). In the particular decay of $B_s$ to $\phi \phi$, the final state 
is a CP eigenstate. Such decays can be used to measure the $B_s$ decay width difference 
($\Delta \Gamma_s$) and the phase responsible for CP violation in the interference 
between decays with and without mixing. 
To conserve the total angular momentum in $B_s \rightarrow \phi \phi$ decays, the 
relative orbital angular momentum between the two $\phi$ mesons in the final state must be either 0, 
1 or 2. In the angular momentum space, there are various bases which can be used to 
analyze decays of pseudo-scalars to two vector mesons, but any formalism involves 
three independent amplitudes for the three different polarizations of the decay products 
in the final state. Measuring the polarization fractions amounts to an important 
test of the corresponding theoretical predictions.  

\begin{figure}[tb]
\centerline{
\includegraphics[width=7cm]{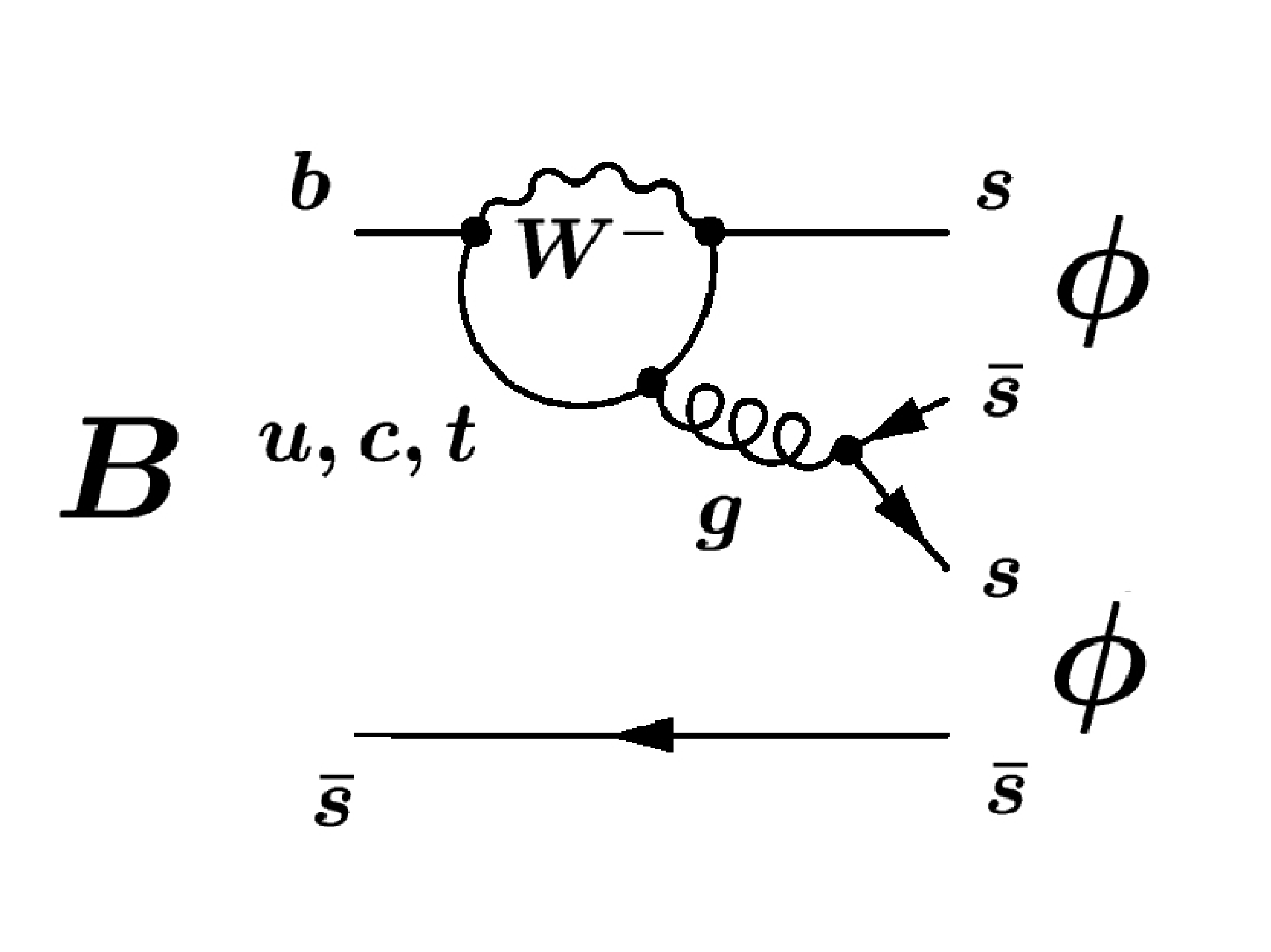}
}
\caption{ Main penguin diagram that contributes to the $B_s \rightarrow \phi \phi$ decay. }
\label{penguin}
\end{figure}

Within the SM, the dominant process that contributes to  
the $B_s \rightarrow \phi \phi$ decay is the $b \rightarrow s \bar{s} s$ penguin 
digram shown in figure~\ref{penguin}. The same penguin amplitude appears in other 
$B \rightarrow VV$ processes 
which exhibit significant discrepancies between the measured polarization fractions and 
the SM predictions. Explanations involving both new physics scenarios as well as 
newly accounted SM effects have been suggested to explain the observations. However, none of the 
existing scenarios is convincing enough. To solve this ``polarization puzzle'' it is 
important to study as many $B \rightarrow VV$ decays as available. 
The first polarization analysis of $B_s \rightarrow \phi \phi$ decays, performed by the 
CDF experiment is presented here together with an updated measurement of the 
$B_s \rightarrow \phi \phi$ branching fraction.

\section{Measurement of the $B_s \rightarrow \phi \phi$ Branching Ratio}

The $B_s \rightarrow \phi \phi$ decay was first observed by the CDF experiment 
in 2005~\cite{Bs_to_phiphi_180pb} using a data sample corresponding to 180~pb$^{-1}$ of integrated 
luminosity. The first measurement of the branching ratio ${\cal B}(B_s \rightarrow \phi \phi)$ 
was performed with 8 signal events and found to be $[1.4 \pm 0.6 (stat.) \pm 0.6 (syst.)] \times 10^{-5}$.  
The analysis was updated in 2009 with a data sample corresponding to 2.9~fb$^{-1}$ of integrated 
luminosity. The data were collected by a trigger which requires two tracks displaced with respect 
to the primary vertex to enhance the contribution from long lived $B$ mesons and suppress 
backgrounds. From the same dataset, $B_s \rightarrow J/\psi \phi$ decay are reconstructed as well 
and used as a normalization mode. This normalization mode was chosen because it has a topology 
similar to the  $B_s \rightarrow \phi \phi$ decay and so, the measured branching ratio will be free 
of uncertainties from $B_s$ and $B_d$ production cross sections, as it would not be using a similar 
$B_d$ penguin decay (e. g. $B_d \rightarrow \phi K^*$).

The $B_s \rightarrow \phi \phi$ decays are reconstructed from to two $\phi(1020)$ vector mesons where 
each $\phi$ meson is reconstructed from the decay $\phi \rightarrow K^{+} K^{-}$. Similarly, 
$B_s \rightarrow J/\psi \phi$ decays are reconstructed from a $J/\psi$ and a $\phi$ meson, where 
the $J/\psi$ decays to $\mu^{+}\mu^{-}$ and the $\phi$ meson decays to $K^{+} K^{-}$. Both $B_s$ decays 
described above lead to four particles in the final state and all four particles come from one 
potentially displaced vertex. The $B_s \rightarrow J/\psi \phi$ is important on its own because 
it may improve the measurement of CP violation previously performed with a sample collected 
with a di-muon trigger~\cite{CDF_beta_s}. The displaced track trigger may add about 25\%  
more $B_s \rightarrow J/\psi \phi$ events which are unique to this independent dataset.   

The events are selected according to an optimization procedure designed to maximize the ratio 
$S/\sqrt(S+B)$, where $S$ is the number of signal events and $B$ is the number of background 
events under the mass signal peak. 
This figure of merit ensures minimal statistical uncertainty 
on the branching ratio measurement and it was verified to also optimize the
uncertainty on the polarization fraction measurement described in the
following section. 
The signal events are simulated while the background events 
are chosen from the $B_s$ mass sidebands. The variables used for the signal selection are chosen 
based on their discriminating power between signal and background. They are verified to be 
un-correlated and to exhibit good agreement between data and simulation. The most important 
variables used for the selection of both $B_s \rightarrow \phi \phi$ and $B_s \rightarrow J/\psi \phi$ 
decays are the transverse decay length of the $B$ vertex projected along the $B$ transverse 
momentum, the impact parameter of the $B$ meson, the quality of the four-track vertex fit 
and transverse momenta of final state particles. In particular, for the $B_s \rightarrow J/\psi \phi$ 
decays, one of the two muons from $J/\psi$ is required to be identified by the CDF muon systems.     

Apart from the combinatorial background which is suppressed by the optimization procedure 
described above, other physics backgrounds are present in this analysis. These physics 
backgrounds come from real $B$ decays which are misreconstructed as either $B_s \rightarrow \phi \phi$ 
or $B_s \rightarrow J/\psi \phi$. In the case of $B_s \rightarrow J/\psi \phi$ decays, the 
main background is $B^0 \rightarrow J/\psi K^{*0}$, where $K^{*0} \rightarrow K^{+} \pi^{-}$. 
When the pion from $K^{*0}$ decay is identified as a kaon, the misreconstructed $K^{*0}$ 
falls in the $\phi$ mass region. The background fraction 
$f_{J/\psi K^{*0}} = N(B^0 \rightarrow J/\psi K^{*0} ) / N(B_s \rightarrow J/\psi \phi)$ is 
estimated using:

\begin{equation}
 f_{J/\psi K^{*0}} = \frac{f_d}{f_s} \frac{ {\cal B}(B^0 \rightarrow J/\psi K^{*0}) }{ {\cal B}(B_s \rightarrow J/\psi \phi) } \frac{{\cal B}(K^{*0} \rightarrow K^+ \pi^- ) }{ {\cal B}(\phi \rightarrow K^+K^-) } \frac{\epsilon^{J/\psi K^{*0}}(J/\psi\phi)}{\epsilon^{J/\psi\phi}}
\end{equation}

where $\epsilon^{J/\psi K^{*0}}(J/\psi\phi)$ is the trigger and selection efficiency of the $B^0 \rightarrow J/\psi K^{*0}$ 
decay reconstructed as $B_s \rightarrow J/\psi \phi$ and $\epsilon^{J/\psi\phi}$ is the trigger and selection efficiency 
for $\epsilon^{J/\psi\phi}$, both determined using simulation. $f_d$ and $f_s$ are the production fractions of the 
$B_d$ and $B_s$ mesons. The fraction $f_{J/\psi K^{*0}}$ is found to be $0.0419 \pm 0.0093$.
For the $B_s \rightarrow \phi \phi$ mode, the physics backgrounds come from $B^0 \rightarrow \phi K^{*0} \rightarrow K^+K^-K^+\pi^-$ 
and $B_s \rightarrow {\bar K}^{*0}K^{*0} \rightarrow K^-\pi^+K^+\pi^- $. Using methods similar to equation 2.1 
(see equations 2 and 3 in~\cite{CDF_pol})  
we find that the contribution of the $B_s \rightarrow {\bar K}^{*0}K^{*0}$ is negligible 
and the contribution of the 
$B^0 \rightarrow \phi K^{*0} \rightarrow K^+K^-K^+\pi^-$ mode is about eight events. 

\begin{figure}[tb]
\centerline{
\includegraphics[width=7cm]{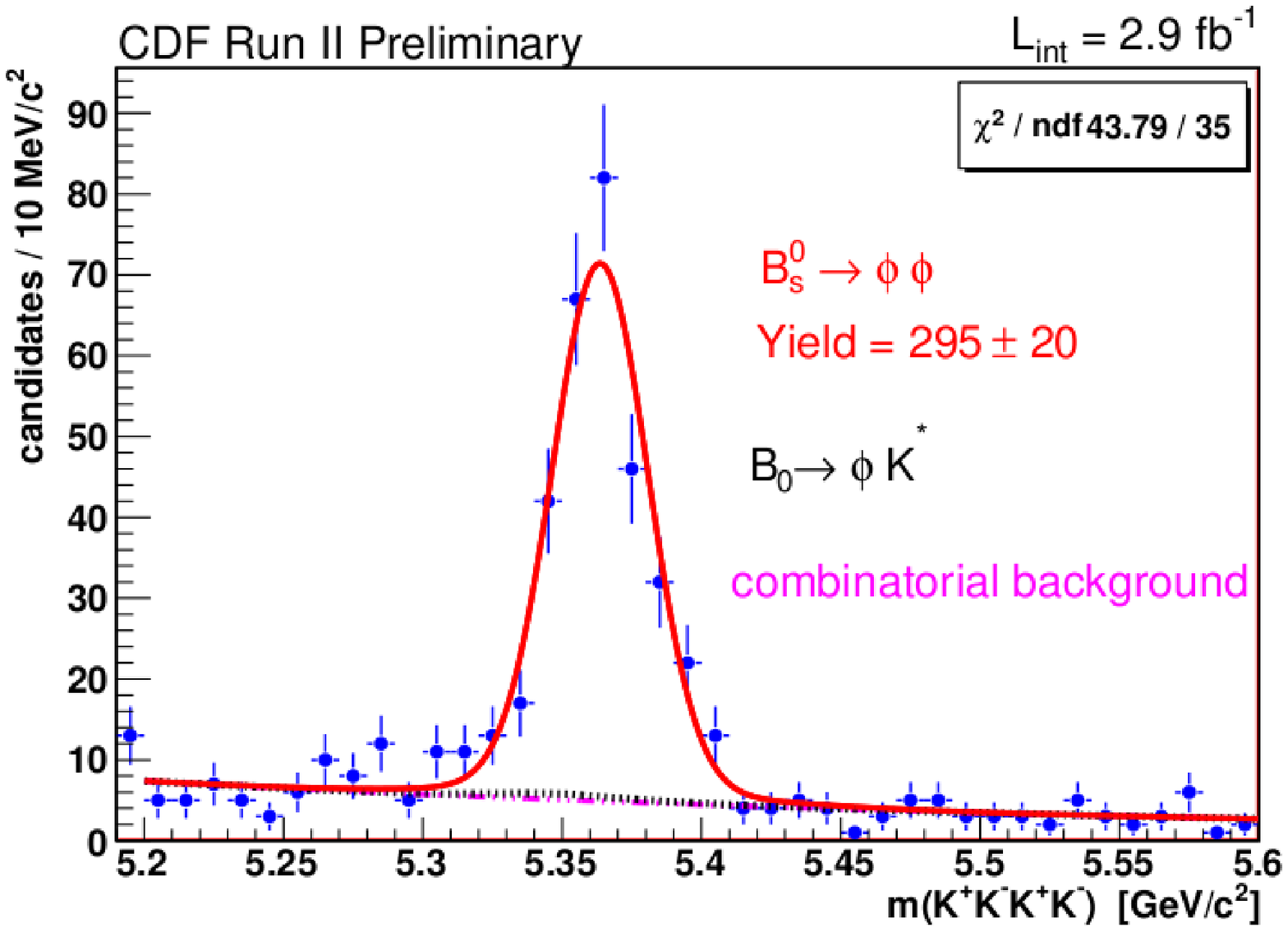}
\includegraphics[width=7cm]{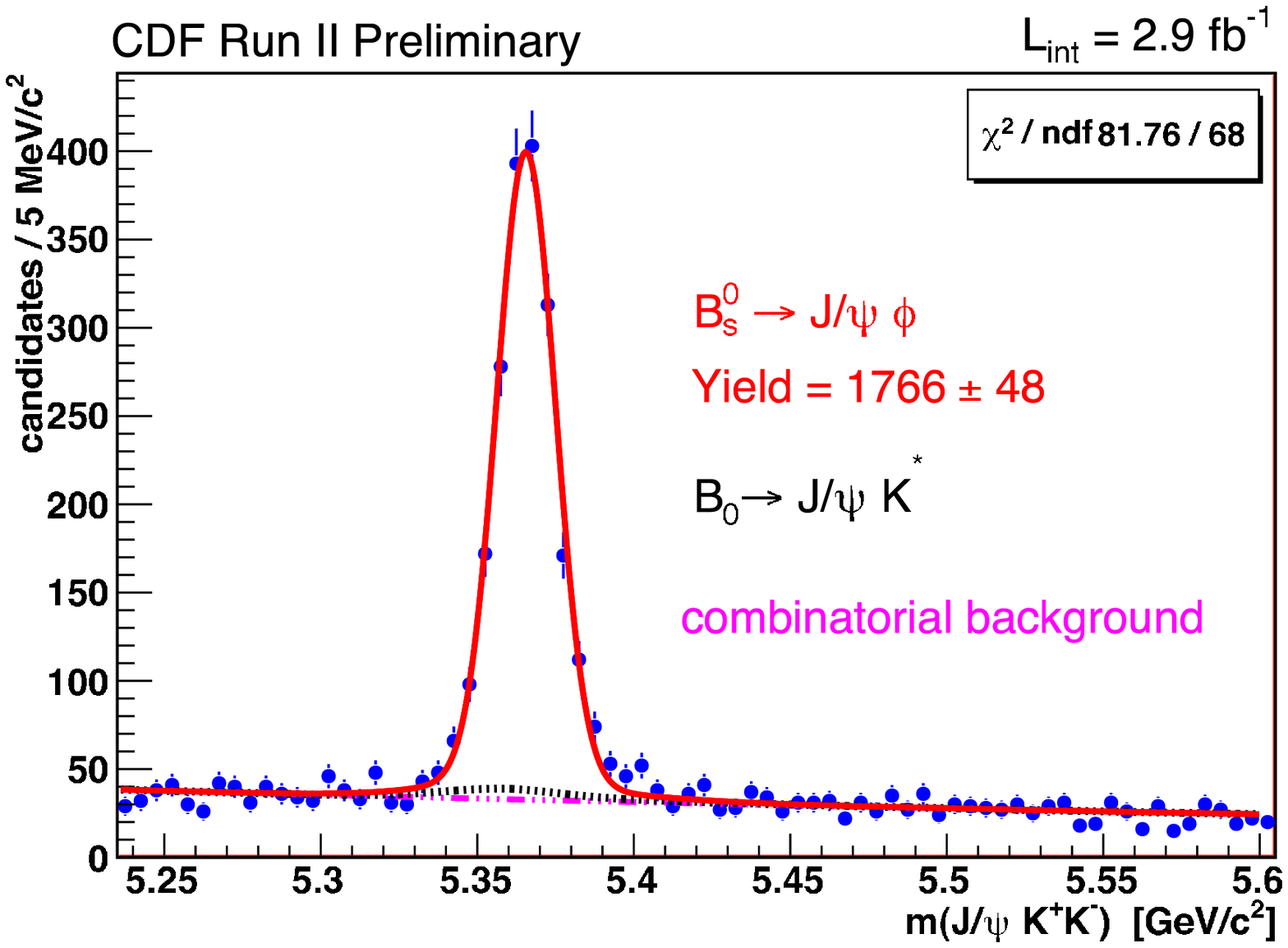}
}
\caption{ Left: $\phi \phi$ invariant mass. Right: $J/\psi \phi$ invariant mass. 
	For both mass distributions physics background contributions are shown 
	together with overlaid fits to the data. }
\label{fig:mass_phiphi_jpsiphi}
\end{figure}

An important step in this analysis is to measure the signal yields of both $B_s \rightarrow J/\psi \phi$
and $B_s \rightarrow \phi \phi$. After applying the optimization procedure described above, the corresponding 
$B_s$ mass peaks are shown in figure~\ref{fig:mass_phiphi_jpsiphi}.
We find $1766 \pm 48 (stat.)$ $B_s \rightarrow J/\psi \phi$ signal events and $295 \pm 20 (stat.)$ 
$B_s \rightarrow J/\psi \phi$ signal events. 

The branching ratio $\cal B$ of the decay $B_s \rightarrow \phi \phi$ normalized to 
the well known ${\cal B}(B_s \rightarrow J/\psi \phi)$ can be evaluated using the 
following equation:

\begin{equation}
   \frac{{\cal B}(B_s \rightarrow \phi \phi)}{{\cal B}(B_s \rightarrow J/\psi \phi)} = \frac{N_{\phi\phi}}{N_{J/\psi\phi}} \times \frac{{\cal B}(J/\psi \rightarrow \mu \mu)}{{\cal B}(\phi \rightarrow KK)} \times \frac{\epsilon_{TOT}^{J/\psi\phi}}{\epsilon_{TOT}^{\phi\phi}} \times \epsilon_{mu}^{TOT}
\end{equation} 

where $N_{J/\psi\phi}$ are $N_{\phi\phi}$ the numbers of $B_s \rightarrow J/\psi \phi$ and 
$B_s \rightarrow \phi \phi$ signal events. $\epsilon_{TOT}^{J/\psi\phi}$ and $\epsilon_{TOT}^{\phi\phi}$ 
are the combined trigger and selection efficiencies. The term $\epsilon_{mu}^{TOT}$ accounts for the 
efficiency of identifying at least one of the muons in the muon detectors.  
Using the above ratio, the uncertainties in the production cross section of the $B$ mesons cancel 
out and several systematic effects due to detector and trigger efficiencies cancel as well, allowing a reduced 
systematic uncertainty in the measurement of the branching ratio. 

The efficiencies for both $B_s \rightarrow \phi \phi$ and $B_s \rightarrow J/\psi \phi$ channels are obtained 
by taking the ratio between the number of simulated events that satisfy the trigger and selection criteria 
and the total number of generated events. The efficiency for the muon identification is determined in a different 
way than the trigger and reconstruction efficiencies because the simulation does not account properly for 
muon acceptance and the corresponding uncertainties would not cancel in the ratio of efficiencies. The muon 
efficiency is determined as a function of the muon momentum and it is obtained by using inclusive $J/\psi \rightarrow \mu \mu$ 
decays reconstructed in the same dataset where either one or both muons have been identified by the muon detectors.    

The most important systematic uncertainties in this analysis are listed here. 
The uncertainties in the number of signal events due to variations in the fit mass range that account 
for the possible presence of unidentified peaking background near the signal peak and uncertainty in the 
shape of the combinatorial background as well as the parameterization of the signal mass 
peak with a single Gaussian function instead of two Gaussians, uncertainties on the physics backgrounds coming 
from errors on the corresponding branching ratios, uncertainties on the muon efficiency, uncertainty 
on the ratio of the trigger and selection efficiencies due to poor knowledge of the polarization amplitudes 
and the decay width difference between the $B_s$ mass eigenstates. The total systematic uncertainty, 
excluding the error on the $B_s \rightarrow J/\psi \phi$ branching ratio is 11\%. 
The final ratio of branching fractions is:

\begin{equation}
   \frac{{\cal B}(B_s \rightarrow \phi\phi)}{{\cal B}(B_s \rightarrow J/\psi \phi)} = [1.78 \pm 0.14 (stat.) \pm 0.20 (syst.)] \times 10^{-2}
\end{equation}   

Using the experimental value of the $B_s \rightarrow J/\psi \phi$ branching ratio we obtain:

\begin{equation}
   {\cal B}(B_s \rightarrow \phi\phi) = [2.40 \pm 0.21(stat.) \pm 0.27 (syst.) \pm 0.82 (BR)] \times 10^{-5}
\end{equation}  
 
where the last uncertainty ($BR$) is the dominant contribution and comes from the error on the 
$B_s \rightarrow J/\psi \phi$ branching ratio.
We note that the world average for the ${\cal B}(B_s \rightarrow J/\psi \phi) = (0.93 \pm 0.33) \times 10^{-3}$ 
is based on a single CDF Run I measurement that assumed the ratio between the $B_s^0$ and $B_d^0$ fragmentation 
fractions $f_s/f_d = 0.40$. The central value of ${\cal B}(B_s \rightarrow J/\psi \phi)$ is scaled to reflect 
the current value of $f_s/f_d = 0.110/0.399 = 0.28$~\cite{PDG}. Consequently, we use 
${\cal B}(B_s \rightarrow J/\psi \phi) = (1.35 \pm 0.46) \times 10^{-3}$.

This result is compatible with the initial observation~\cite{Bs_to_phiphi_180pb}, with substantial 
improvement on the statistical uncertainty. The result is also compatible with recent theoretical 
calculations~\cite{TH_br1} and~\cite{TH_br2}.

\section{Measurement of the Polarization Amplitudes in $B_s \rightarrow \phi \phi$ Decays}

As already pointed out in the Introduction, in the $B_s \rightarrow \phi \phi$ decay, 
the dominant diagram is the $b \rightarrow s$ penguin shown in figure~\ref{penguin}. 
The same penguin amplitude is also relevant in in other processes which have shown deviations 
from the SM predictions. Such effects are the difference in the CP asymmetries in $B_d \rightarrow K^+ \pi^-$ 
and $B^+ \rightarrow K^+ \pi^0$ and the potential difference between the sin(2$\beta$) measurements 
in $b \rightarrow s {\bar q} q$ and $b \rightarrow c {\bar c} s$ $B^0$ decays. 

The decay amplitude in $B_s \rightarrow \phi \phi$ decays can be expressed in terms of three independent 
decay amplitudes, which correspond to the three possible relative angular momenta between 
the two $\phi$ vector mesons. In this analysis we use the helicity formalism 
in which the polarizations 
of the two vector mesons are either longitudinal with respect to the direction of motion $A_0$ 
or transverse relative to the direction of motion. 
There are two transverse amplitudes $A_{||}$ and 
$A_{\perp}$ corresponding to the two polarizations being parallel or perpendicular to each other. 
The fractions of these amplitudes can be measured from the analysis of the angular distributions 
of the final state particles (the decay products of the two $\phi$ mesons).  

Taking into account the V-A nature of the weak interaction and the helicity conservation in QCD, 
it is expected that the dominant amplitude is the longitudinal polarization while the transverse 
component is suppressed by a factor of $m_V/m_B$~\cite{TH_br2}. This expectation is confirmed in 
tree-level dominated $b \rightarrow u$ transitions like $B^0 \rightarrow \rho^+ \rho^-$~\cite{ct3, ct4}, 
$B^+ \rightarrow \rho^0, \rho^+$~\cite{ct5} and $B^+ \rightarrow \omega \rho^+$~\cite{ct6}, but it 
is not confirmed in $B \rightarrow \phi K^*$, a ${\bar b} \rightarrow {\bar s}$ decay. In this decay, 
the transverse polarization fraction is about equal to the longitudinal polarization~\cite{ct8, ct9, ct10}. 
This unexpected result is known as the ``polarization puzzle''. Explanations involving either new 
physics~\cite{ct11, TH_br1} or SM corrections including either penguin annihilation~\cite{TH_br1, TH_br2, ct14} 
or final state interactions~\cite{ct15, ct16, ct17, ct18} have been proposed.  
Recent work~\cite{HaiYangCheng_new} based on QCD predictions finds the longitudinal polarization 
fraction to be in excellent agreement with our measurement if the penguin
annihilation amplitude is fitted to the $B \rightarrow \phi K^{*}$ data.  

We present the first measurement of the polarization fractions in $B_s \rightarrow \phi \phi$ decays 
using 2.9~fb$^{-1}$ of CDF data. As a cross check, we also perform the polarization fractions in  
$B_s \rightarrow J/\psi \phi$ decays. Both measurements are performed on the data samples 
selected for the branching ratio measurement described in section 2. For either of the two decays, 
we refer to the two vector mesons as $V_1$ and $V_2$ and to their decay products as final state particles 
$P_1$ and $P_2$ from $V_1$ and $P_3$ and $P_4$ from $V_2$.  

We use the helicity formalism to describe $B_s \rightarrow \phi \phi$ decays. The $x'$ and $x''$ axes are defined 
as the directions of the $V_1$ and $V_2$ momenta in the rest frame of the $B_s$ meson, respectively. We define 
the angle $\theta_1$ ($\theta_2$) as the angle between the $x'$ ($x''$) axis and the $P_1$ ($P_3$) momentum vector, 
defined in the rest frame of $V_1$ ($V_2$). The $\Phi$ angle is defined as the angle between the decay planes 
of the two daughter particles. The three angles $\vec{\omega} = (\theta_1, \theta_2, \Phi)$ completely describe the directions 
of the final state particles. The distributions of these angles are used to separate the three amplitudes 
and determine their corresponding fractions. The probability distribution function (PDF) used to describe 
the helicity angular distribution for the signal is obtained by integrating out the time dependence. 
The differential decay rate as function of the helicity angles is given by:

\begin{equation}
 \frac{d^3\Lambda(\vec{\omega})}{d\vec{\omega}} = \frac{9}{32\pi} \frac{1}{W}[F_e(\vec{\omega}) + F_o(\vec{\omega})]
\end{equation}  

where 
$F_e = \frac{2}{\Gamma_L}[|A_0|^2 f_1(\vec{\omega}) + |A_{||}|^2 f_2(\vec{\omega}) + |A_0||A_{||}| cos(\delta_{||}) f_5(\vec{\omega}) ]$, 
$F_o = \frac{2}{\Gamma_H} |A_{\perp}|^2 f_3(\vec{\omega})$, $W = \frac{|A_0|^2 + |A_{||}|^2}{\Gamma_L} + \frac{|A_{\perp}|^2}{\Gamma_H}$. 
Here, $\Gamma_L$ and $\Gamma_H$ are the decays widths of the $B_s$ mass eigenstates, $f_i$ are functions of the 
helicity angles $\vec{\omega}$ and $\delta_{||}$ is a strong phase defined as $\delta_{||} = arg(A^*_0 A_{||})$.
The decay widths $\Gamma_L$ and $\Gamma_H$ are fixed to the
world average. Although our trigger gives a non flat
acceptance as a function of $B_s$ proper decay time, this time-integrated
approach has been verified to give biases smaller that the statistical
uncertainty of the polarization fraction measurement by using simulation 
and the measurement of the equivalent fractions in the $J/\psi \phi$ control sample.
A similar formalism is used to describe $B_s \rightarrow J/\psi \phi$ decays in the transversity basis~\cite{transversity_basis}. 
The data samples and the optimization procedures are the same as the ones in the branching ratio measurement described in section 2.

\begin{figure}[tb]
\centerline{
\includegraphics[width=5cm]{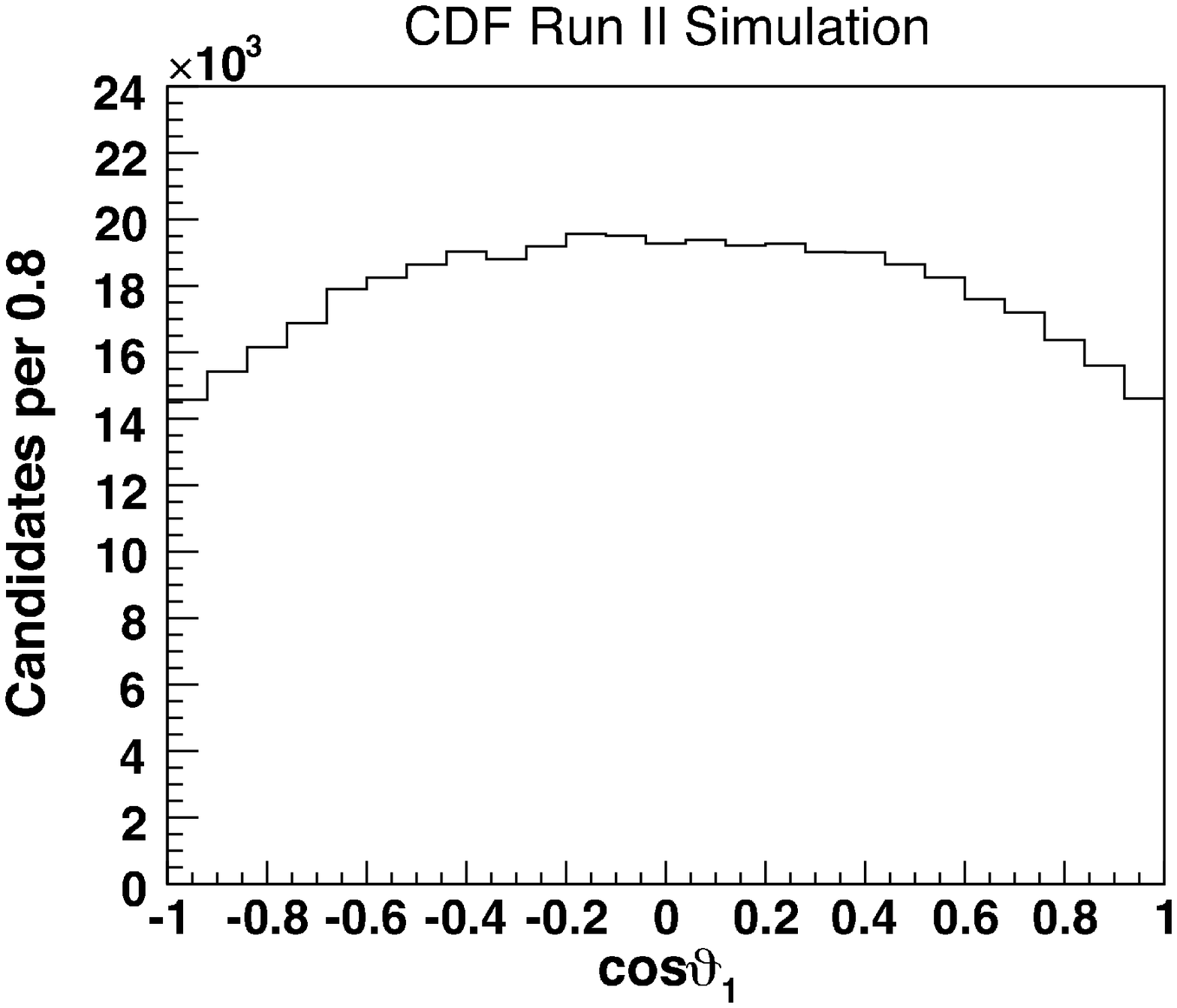}
\includegraphics[width=5cm]{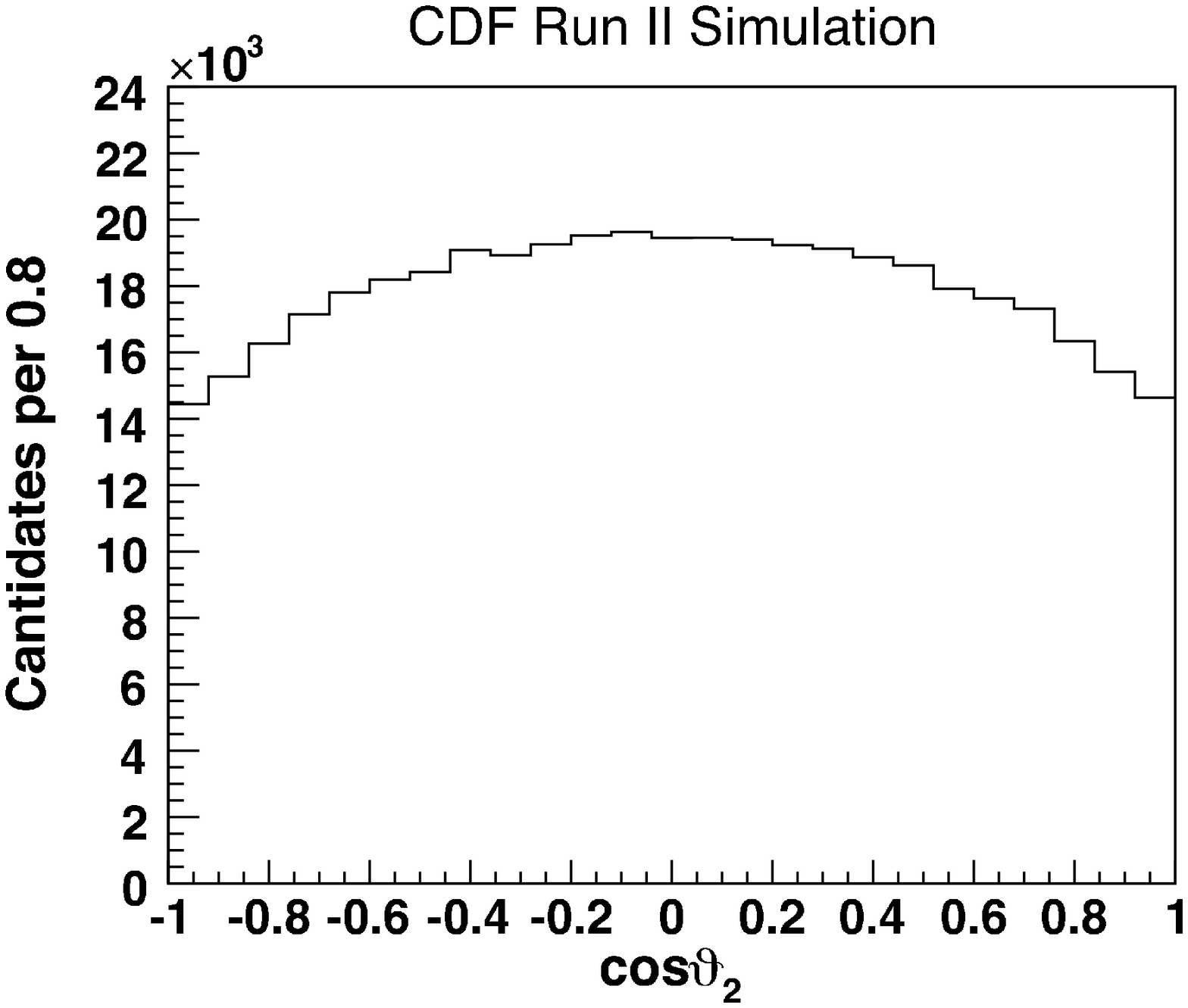}
\includegraphics[width=5cm]{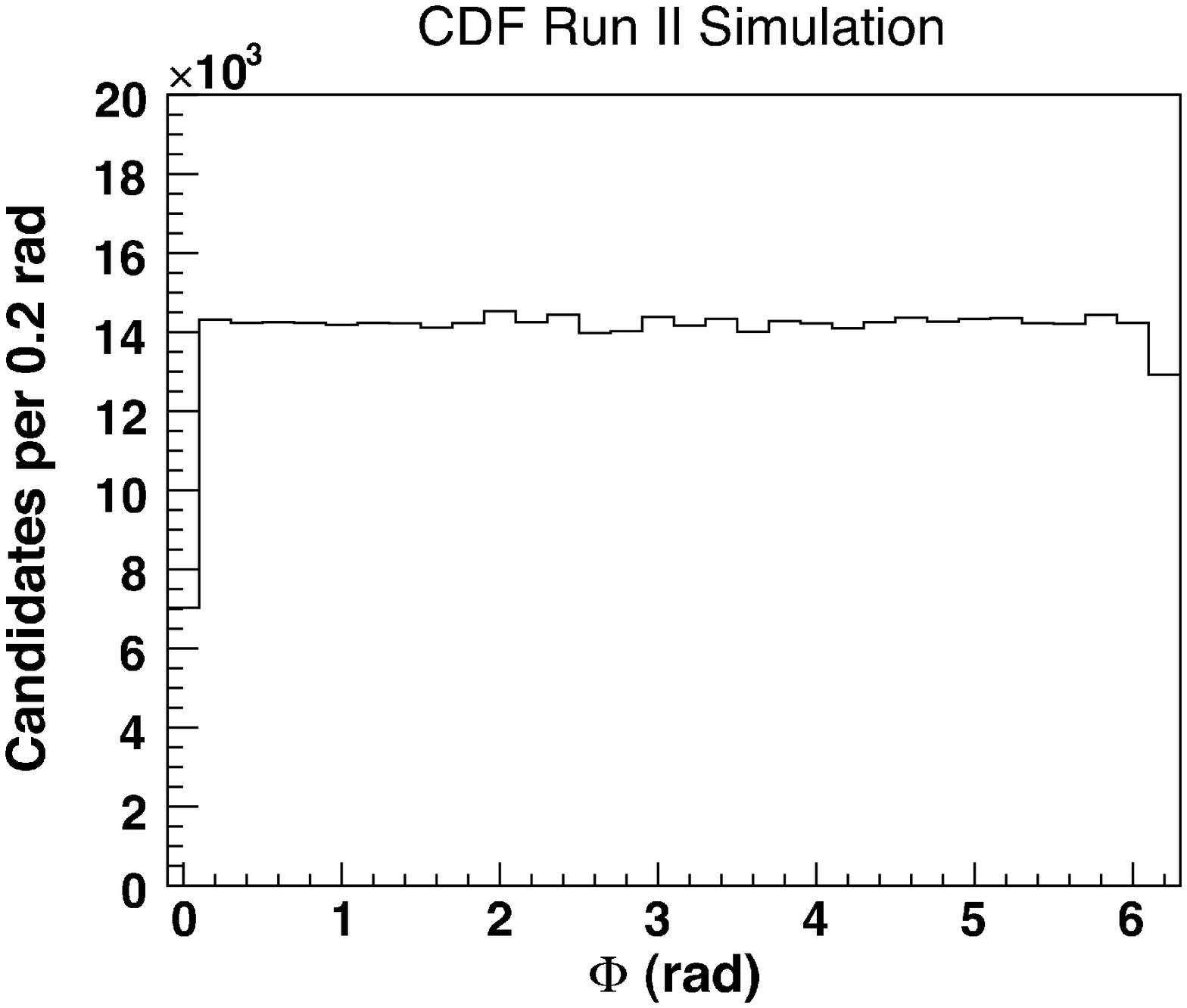}
}
\caption{ Detector angular acceptance projections for the helicity angles cos($\theta_1$), cos($\theta_2$) and $\Phi$ 
	used for the angular analysis of $B_s \rightarrow \phi \phi$ decays. }
\label{fig:angular_acceptance}
\end{figure}

The observables measured in this analysis are the polarization fractions $|A_0|^2$ and $|A_{||}|^2$ as well as the relative strong 
phase between them $\delta_{||}$. The measurement of these observables is performed using an unbinned maximum likelihood 
fit using as event-by-event inputs the reconstructed mass of the $B_s$ candidate and the reconstructed helicity angles.    
The mass distribution is used in the fit to discriminate the signal from background. The angular distributions separate between 
the three polarization amplitudes. The signal mass distribution has a width of $~20$~MeV/c$^2$ for the $B_s \rightarrow \phi \phi$
and $10$~MeV/c$^2$ for the $B_s \rightarrow J/\psi \phi$. In both cases the signal is parameterized with two Gaussian functions 
with the same mean and different resolutions. The mass background distributions are described by exponential functions. 
The PDFs used to describe the helicity angular distributions for the signal are 
described in~\cite{CDF_pol}. 
The observed angular distributions in both helicity and transversity bases are different than the expected theoretical 
distributions due to detector acceptance effects. The angular acceptance is determined using simulated signal events. 
The projections on the helicity angles $\theta_1$, $\theta_2$ and $\Phi$ are shown in figure~\ref{fig:angular_acceptance}.  
The background angular distributions are determined from the $B_s$ mass sidebands. These distributions are parameterized 
with empirical functions. The $\Phi$ distribution is parameterized with a constant function and the angles $\theta_1$ 
and $\theta_2$ are parameterized with functions of the form $1 + B \times $cos$^2(\theta)$ where $B$ is a parameter 
determined by the fit. 
Before performing the measurement of the polarization fractions in $B_s \rightarrow \phi \phi$, several tests 
of the unbinned maximum likelihood fit are performed. The fit is tested on pseudo-experiments where no biases 
are found and the uncertainties are in the Gaussian regime. The polarization fractions are measured in 
$B_s \rightarrow J/\psi \phi$ decays used as a control sample:
\begin{equation}
 |A_0|^2 = 0.534 \pm 0.019 (stat.),  \ \ \  
 |A_{||}| = 0.220 \pm 0.025 (stat.).  
\end{equation}
 In this case the polarization fractions 
are found to be in good agreement with previous CDF measurements from a di-muon sample~\cite{betas}. Finally, samples of 
$B_s \rightarrow \phi \phi$ 
are generated and passed through the full trigger and detector simulation and then through the analysis 
selection. The polarizations are measured in these samples and good agreement with the generated values is 
found.

\begin{figure}[tb]
\centerline{
\includegraphics[width=7cm]{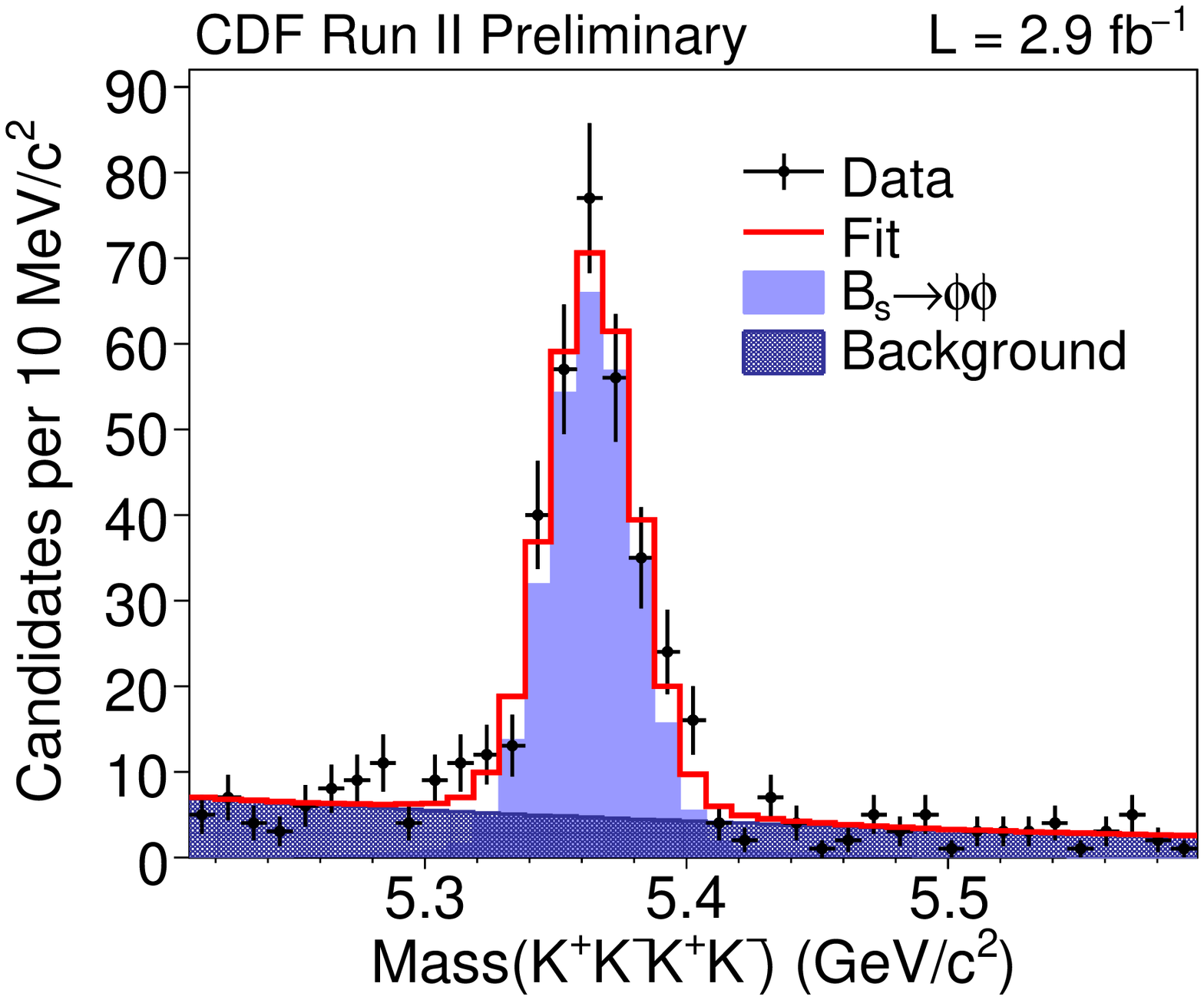}
\includegraphics[width=7cm]{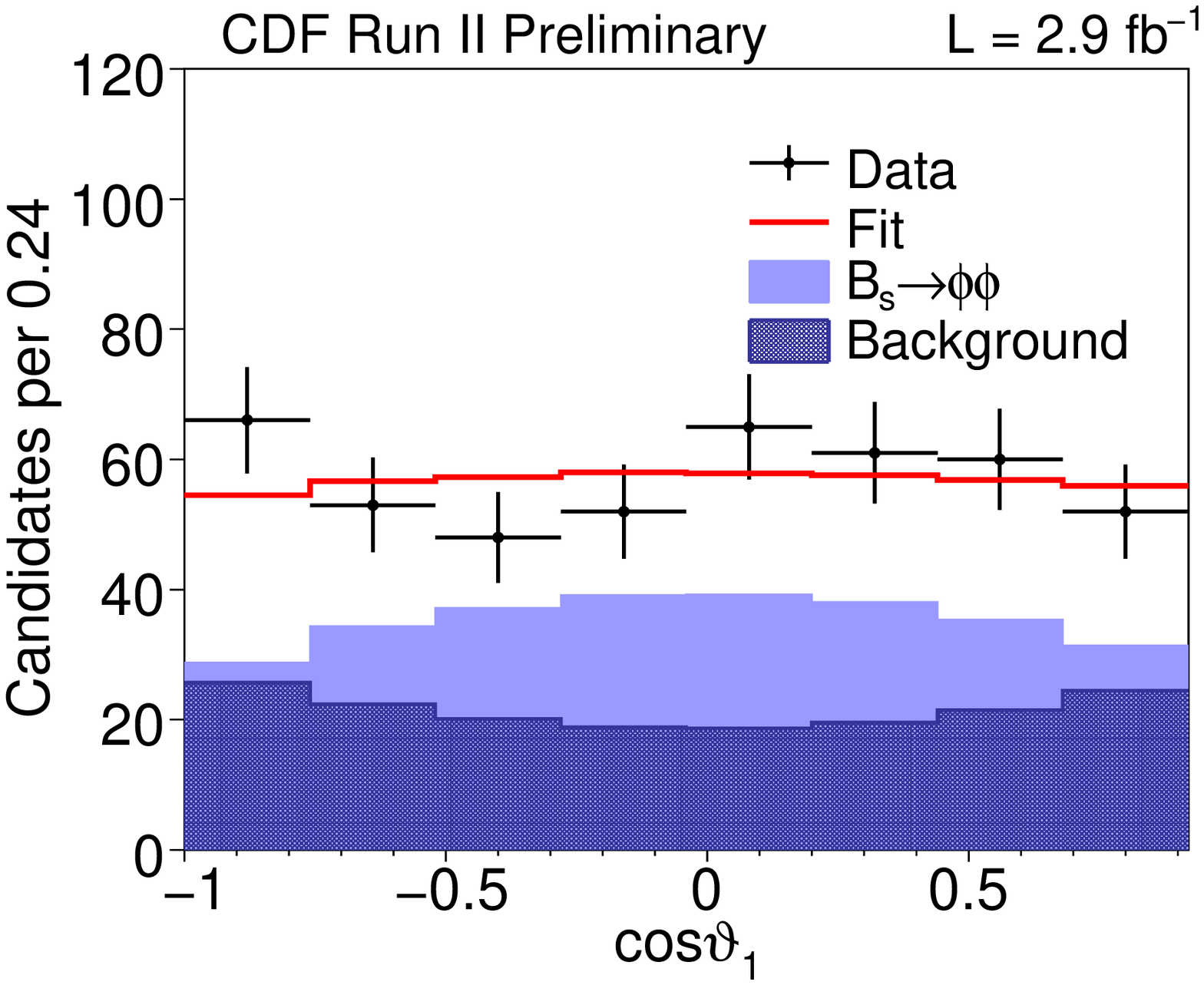} 
}
\centerline{
\includegraphics[width=7cm]{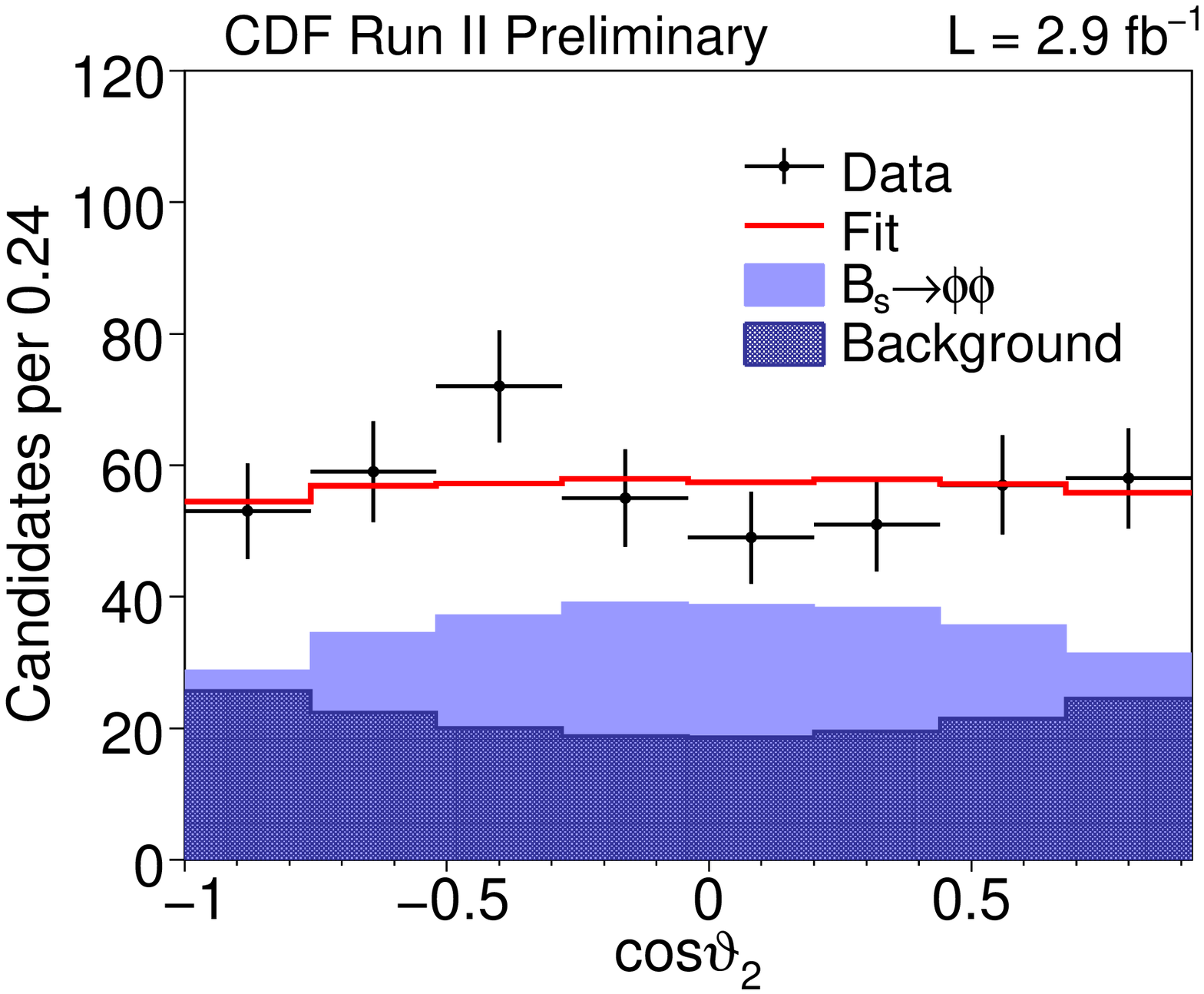}
\includegraphics[width=7cm]{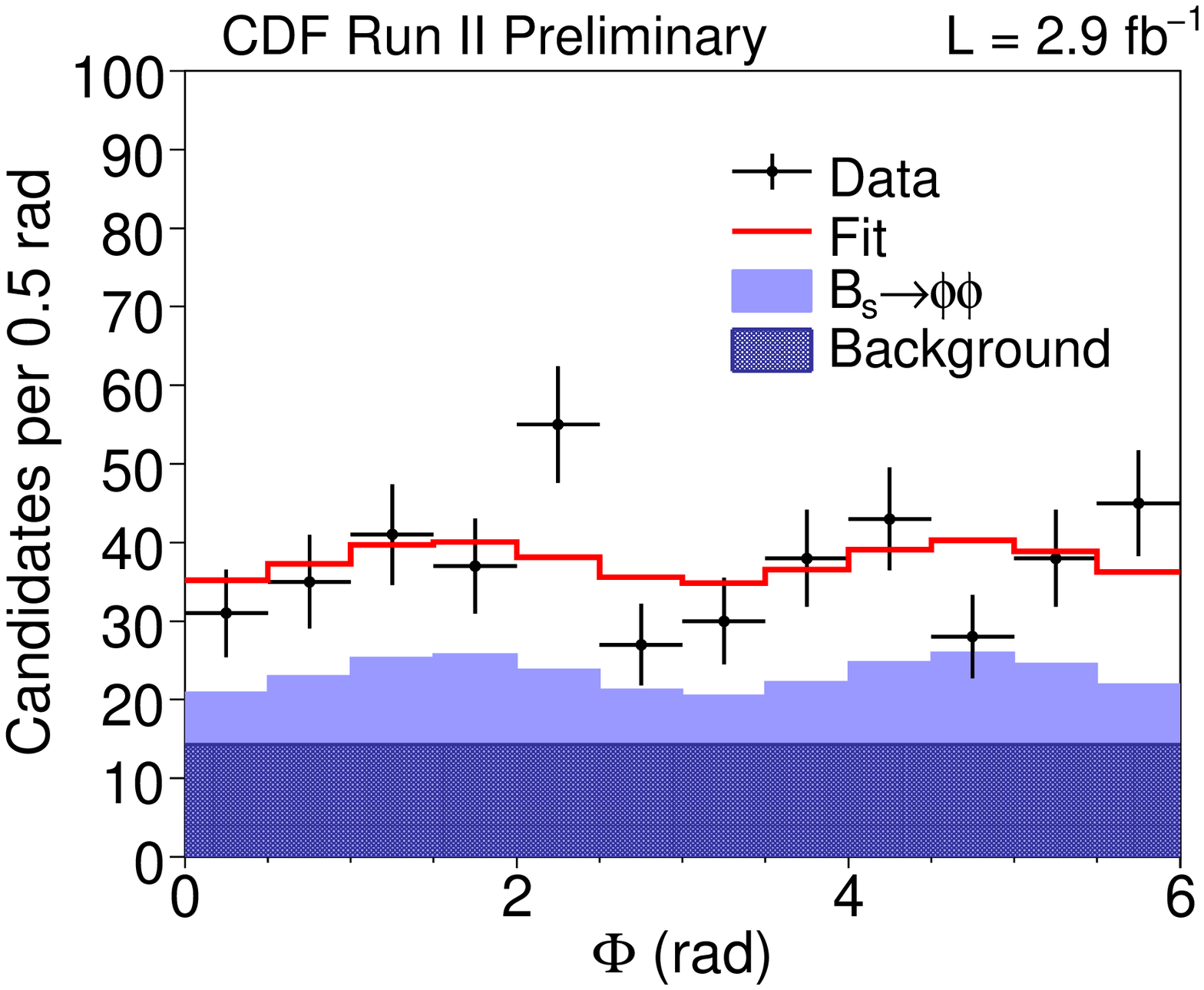}
}
\caption{ Fit projections for the mass component and the angular components in  $B_s \rightarrow \phi \phi$ decays.  }
\label{fig:helicity_projections}
\end{figure}

Finally, we measure the polarization fractions in $B_s \rightarrow \phi \phi$ decays:
\begin{equation}
 |A_0|^2 = 0.348 \pm 0.041 (stat.),  \ \ \  
 |A_{||}| = 0.287 \pm 0.043 (stat.).  
\end{equation}
The measured strong phase is $cos(\delta_{||}) = -0.91^{+0.15}_{-0.13}$.
The fit projections onto the mass and helicity angles are shown in figure~\ref{fig:helicity_projections} 
which shows very good agreement between the data distributions and the fitting functions.  

The main systematic uncertainties on the $B_s \rightarrow \phi \phi$ polarization fractions 
come from the dependence of the angular acceptance on the decay width 
difference $\Delta \Gamma_s$, uncertainties on the lifetimes of the heavy and light $B_s$ 
mass eigenstates $\tau_H$ and $\tau_L$ and the potential $KK$ s-wave contributions to the 
angular distributions.   

The final results, including systematic uncertainties are:

\begin{equation}
 |A_0|^2 = 0.348 \pm 0.041 (stat.) \pm 0.021 (syst.)  
\end{equation}
\begin{equation}
 |A_{||}|^2 = 0.287 \pm 0.043 (stat.) \pm 0.011 (syst.)  
\end{equation}
\begin{equation}
 |A_{\perp}|^2 = 0.365 \pm 0.044 (stat.) \pm 0.027 (syst.)  
\end{equation}
\begin{equation}
 cos(\delta_{||}) = -0.91^{+0.15}_{-0.13} (stat.) \pm 0.09 (syst.)   
\end{equation}

The longitudinal and transverse polarization fractions are:
\begin{equation}
 f_L = 0.348 \pm 0.041 (stat.) \pm 0.021 (syst.)  
\end{equation}
\begin{equation}
 f_T = 0.652 \pm 0.041 (stat.) \pm 0.021 (syst.)  
\end{equation}
 
It is clear from this measurement that the SM expected amplitude hierarchy 
$|A_0| \gg |A_{||}| \simeq |A_{\perp}|$ is not valid in $B_s \rightarrow \phi \phi$ 
decays. Instead, the observed relation between the polarization amplitudes 
is given by: $|A_0| \simeq |A_{||}| \gtrsim |A_{\perp}|$, which is similar 
to the measurements for the $\bar b \rightarrow \bar s$ penguin transition 
of $B \rightarrow \phi K^*$ decays~\cite{ct8, ct25, ct26} which were the 
origin of the polarization puzzle.

We compare our results with various theoretical predictions of the polarization 
amplitudes. We find that our central values are consistent within the uncertainty 
ranges with the expectations of the QCD factorization~\cite{TH_br1}, while they are not 
in good agreement with the expectation of perturbative QCD~\cite{TH_br2} and 
QCD factorization~\cite{HaiYangCheng_new}.

\section{Conclusions}

We have presented an updated measurement of the $B_s \rightarrow \phi \phi$ 
branching ratio using a data sample corresponding to 2.9~fb$^{-1}$ of integrated 
luminosity. Using the same data sample, we measured for the first time the polarization 
fractions in $B_s \rightarrow \phi \phi$ decays. The measured amplitudes 
confirm the previously observed polarization puzzle in certain $B \rightarrow VV$ decays. 

Each of the two Tevatron experiments have currently accumulated about 7~fb$^{-1}$ of data  
and expect 10~fb$^{-1}$ by the end of the Tevatron running in 2011. With a sample three 
times as large, CDF will improve the statistical errors on the polarization amplitudes 
in $B_s \rightarrow \phi \phi$ by a factor of two and will attempt to measure the 
decay width difference $\Delta \Gamma_s$ in this mode. Further studies of rare $B_s$ 
decays and CP violation in the $B_s$ will be improved with more data.


\begin{thebibliography}{99}
\bibitem{rare_decays} M. Aoki, Tevatron Results on $B_s \rightarrow \mu \mu$ and $B_s \rightarrow K^* \mu \mu$, FPCP2010 proceedings.
\bibitem{CDF_beta_s} L. Oakes, Measurement of $\beta_s$ at CDF, FPCP2010 proceedings.
\bibitem{D0_beta_s} A. Chandra, Measurement of $\beta_s$ at D0, FPCP2010 proceedings.
\bibitem{D0_di_muon} G. Brooijmans, Evidence for an anomalous like-sign dimuon charge asymmetry, FPCP2010 proceedings.
\bibitem{CDF_br} {\it http://www-cdf.fnal.gov/physics/new/bottom/090618.blessed-Bsphiphi2.9/ }
\bibitem{CDF_pol} {\it http://www-cdf.fnal.gov/physics/new/bottom/100304.blessed-Bsphiphi\_amplitudes/index.html }
\bibitem{Bs_to_phiphi_180pb} D. Acosta {\em et al.} (CDF  Collaboration), Phys. Rev. Lett. {\bf 95}, 031801, 2005 
\bibitem{PDG} Particle Data Group; Phys. Lett. {\bf B667}, 1 (2008)
\bibitem{TH_br1} M. Beneke, J. Rohrer and D. Yang, Nucl Phys. B {\bf 774}, 64 (2007)
\bibitem{TH_br2} A. Ali, G. Kramer, Y. Li {\em et al}, Phys. Rev. D {\bf 76}, 074018 (2007)
\bibitem{ct3} Belle Collaboration, Phys. Rev. Lett., 96:171801, 2006
\bibitem{ct4} B. Aubert {\em et al}, Phys. Rev. D, 76:052007, 2007
\bibitem{ct5} BaBar Collaboration, Phys. Rev. Lett, 97(26):261801, 2006
\bibitem{ct6} BaBar Collaboration, Phys. Rev. D, 74(5):051102, 2006
\bibitem{ct8} BaBar Collaboration (B. Aubert {\em et al}), Phys. Rev. Lett. 98:051801, 2007 
\bibitem{ct9} BaBar Collaboration (B. Aubert {\em et al}), Phys. Rev. Lett. 97(20):201801, 2006
\bibitem{ct10} K. F. Chen for the Belle Collaboration, Phys. Rev. Lett., 94:221804, 2005 
\bibitem{ct11} E. Alvarez {\em et al}, Phys. Rev. D, 70:115014, 2004 
\bibitem{ct12} C. S. Kim, Y.-D. Yang, ``Polarization anomaly in $B \rightarrow Phi K^*$ and probe of tensor interactions, arXiv:hep-ph/0412364, 2004
\bibitem{ct14} A. L. Kagan, Phys. Lett., B601:151-163. 2004
\bibitem{ct15} P. Colangelo {\em et al}, Phys. Lett. B597:291-298, 2004
\bibitem{ct16} M. Ladisa {\em et al}, Phys. Rev. D70:114025, 2004
\bibitem{ct17} H.-Y. Cheng {\em et al}, Phys. Rev. D71:014030, 2005
\bibitem{HaiYangCheng_new} H.~Y.~Cheng and C.~K.~Chua,
	Phys. Rev. D80:114026, 2009
\bibitem{ct18} Ch. Bauer {\em et al}, Phys. Rev. D70:054015, 2004
\bibitem{transversity_basis} A.~S. Dighe {\em et al}, Eur. Phys. J. C {\bf 6}, 647 (1999).
\bibitem{betas} CDF Collaboration, T. Aaltonen {\em et al} Phys. Rev. Lett., 100:121803, 2008
\bibitem{ct25} Belle Collaboration, Phys. Rev. Lett., 101(23):231801, 2008 
\bibitem{ct26} BaBar Collaboration, Phys. Rev. Lett., 99(20):201802, 2007

\end{thebibliography}
\end{document}